\documentclass[showpacs,aps,graphicx,twocolumn]{revtex4}%
\usepackage{graphicx}

\begin{document}
\title{Efficient symmetric multiparty quantum state sharing of an arbitrary $m$-qubit state}
\author{ Xi-Han Li,$^{1,2}$ Ping Zhou,$^{1,2}$  Chun-Yan Li,$^{1,2}$  Hong-Yu Zhou,$^{1,2,3}$
and  Fu-Guo Deng$^{1,2,3}$\footnote{ Email: fgdeng@bnu.edu.cn} }
\address{$^1$ The Key Laboratory of Beam Technology and Material
Modification of Ministry of Education, Beijing Normal University,
Beijing 100875,
People's Republic of China\\
$^2$ Institute of Low Energy Nuclear Physics, and Department of
Material Science and Engineering, Beijing Normal University,
Beijing 100875, People's Republic of China\\
$^3$ Beijing Radiation Center, Beijing 100875,  People's Republic
of China}
\date{\today }

\begin{abstract}
We present a scheme for symmetric multiparty quantum state sharing
of an arbitrary $m$-qubit state with $m$
Greenberger-Horne-Zeilinger states following some ideas from the
controlled teleportation [Phys. Rev. A \textbf{72}, 02338 (2005)].
The sender Alice performs $m$ Bell-state measurements on her $2m$
particles and the controllers need only to take some single-photon
product measurements on their photons independently, not
multipartite-entanglement measurements, which makes this scheme
more convenient than the latter. Also it does not require the
parties to perform a controlled-NOT gate on the photons for
reconstructing the unknown $m$-qubit state and it is an optimal
one as its efficiency for qubits approaches 100\% in principle.
\end{abstract}
 \maketitle

\section{introduction}

Suppose a president of a bank, say Alice wants to send some secret
message $M_A$ to her agents, Bob and Charlie who are at a remote
place for her business. She doubts that there may be one of the
two persons dishonest and he will destroy the business with the
message independently. Alice believes that the honest one will
keep the dishonest one from doing any damage if they both appear
in the process of dealing with the business. In this way, the
message cannot be transmitted to the agents directly. In classical
secret sharing \cite{Blakley}, Alice splits $M_A$ into two pieces,
$M_B$ and $M_C$, and sends them to Bob and Charlie, respectively.
When they act in concert, they can recover the message
$M_A=M_B\oplus M_C$; otherwise, they can obtain nothing about the
message. The classical signal is in one of the eigenvectors of a
measuring basis (MB), say $\sigma_z$, and it can be copied fully
and freely \cite{book}. It is impossible for the parties to
communicate in an unconditionally secure way if they only resort
to classical communication \cite{QKD1}. When quantum mechanics
enters the field of information, the story is changed
\cite{QKD1,QKD2}.

Quantum secret sharing (QSS) is the generalization of classical
secret sharing into quantum scenario. Most existing QSS schemes
are focused on creating a private key among several parties or
splitting a classical secret. For example, an original QSS scheme
\cite{HBB99} was proposed by Hillery, Bu\v{z}ek, and Berthiaume
(HBB) in 1999 by using three-particle and four-particle entangled
Greenberger-Horne-Zeilinger (GHZ) states for distributing a
private key among some agents and sharing classical information.
In HBB scheme, the three parties, Alice, Bob and Charlie choose
randomly two measuring bases (MBs), $\sigma_x$ and $\sigma_y$ to
measure their particles independently. The probability that the
quantum resource can be used for carrying the useful information
is 50\%. That is, their results are correlated and will be kept
for creating a key when they all choose the MB $\sigma_x$ or one
chooses $\sigma_x$ and the others choose $\sigma_y$, which takes
place with the probability 50\%. They removed the ideas in the
controlled teleportation \cite{ControlledTele} to split a
classical secret among the agents. Subsequently, Karlsson, Koashi
and Imoto proposed another QSS protocol for those two goals with
multi-particle entangled states and entanglement swapping
\cite{KKI}. Its intrinsic efficiency for qubits $\eta_q$
\cite{Cabello}, the ratio of number of theoretical valid
transmitted qubits to the number of transmitted qubits is about
50\% as half of the instances will be abandoned \cite{deng2005}.
Now, there are some theoretic schemes for sharing and splitting a
classical message
\cite{HBB99,KKI,deng2005,Gottesman,Nascimento,yangcpQSS,Karimipour1,
Tyc,longqss,deng20052,guoqss,zhangPLA,Bandyopadhyay,Karimipour,zhanglm,YanGao}.
The experiment demonstration of QSS was also studied by some
groups \cite{TZG,AMLance}.

Recently, a novel concept, quantum state sharing (QSTS) was
proposed and actively pursued by some groups
\cite{cleve,Peng,dengmQSTS,AMLance}. It is the extension of QSS
for sharing an unknown state among several agents by resorting to
the multipartite entanglements. Cleve, Gottesman and Lo
\cite{cleve} introduced a way for a $(k,n)$ threshold QSTS scheme
which can be used to split a secret quantum state related to a
classical secret message \cite{dengmQSTS}. Li et al. \cite{Peng}
proposed a scheme for sharing an unknown single qubit with some
Einstein-Podolsky-Rosen (EPR) pairs and a multi-particle joint
measurement. In 2004, Lance et al. studied QSTS with continuous
variable \cite{AMLance}. Deng et al. \cite{dengmQSTS} introduced a
scheme for splitting an arbitrary two-qubit state with EPR pairs
and GHZ-state measurements. Also, some controlled teleportation
schemes were discussed in
Ref.\cite{ControlledTele,ControlledTele2,ControlledTele3}. The
unknown state can be teleported to the sender with the control of
some controllers. In fact, almost all those controlled
teleportation schemes can be used for QSTS with or without a
little modification. For example, in Ref. \cite{ControlledTele3},
QSTS for an arbitrary two-qubit state can be implemented with the
symmetric controlled teleportation scheme \cite{ControlledTele3}
without any modification. In essence, a secure QSTS scheme can
also be used for controlled teleportation by means that some
agents act as the controllers in the latter.

In this paper, we will present a scheme for QSTS of an arbitrary
$m$-particle state following some ideas in Ref.
\cite{ControlledTele3}. It will be shown that the agents need only
to perform a product measurement $\sigma^1_x \otimes \sigma^2_x
\otimes \ldots \otimes\sigma^m_x$ on their particles, not
multipartite-entanglement measurements, which makes this scheme
more convenient than Ref. \cite{ControlledTele3}. Moreover, it
requires the agents only to perform some local unitary operations
on the photons retained for recovering the unknown state, not a
two-qubit joint operation, such as controlled-NOT operation.

\section{QSTS of an arbitrary $M$-qubit state}

\subsection{QSTS of an arbitrary two-qubit state with two agents}

For simplicity, we first present a way for the symmetric quantum
state sharing of an arbitrary two-qubit state with two
three-particle GHZ states. That is, there are two agents, Bob and
Charlie. We assume that the parties share a sequence of
multipartite entangled states securely first, similarly as Refs.
\cite{two-step,ControlledTele3}. Similar to quantum secure direct
communication \cite{two-step,QSDC}, the parties can do quantum
privacy amplification on the sequence of entangled states shared
with quantum entanglement purification \cite{qpap1,qpap2} in a
noise channel.

The basic idea of this QSTS scheme can be described as following.
Suppose that the quantum information is an unknown arbitrary
two-qubit state
\begin{eqnarray}
\vert \Phi\rangle_{xy}=a\vert 00\rangle_{xy} + b \vert
01\rangle_{xy} + c\vert 10\rangle_{xy} + d\vert
11\rangle_{xy},\label{unknownstate}
\end{eqnarray}
where $\vert 0\rangle$ and $\vert 1\rangle$ are the two
eigenvector of the MB $\sigma_z$ (for example the polarization of
the single photon along the z-direction), and
\begin{eqnarray}
\vert a\vert^2 + \vert b\vert^2 + \vert c\vert^2 + \vert d\vert^2=1.
\end{eqnarray}
Alice prepares two three-particle GHZ states
\begin{eqnarray}
\left\vert\Psi\right\rangle_{a_1a_2a_3}=\left\vert\Psi\right\rangle_{b_1b_2b_3}
=\frac{1}{\sqrt{2}}(\left\vert 000\right\rangle+\left\vert
111\right\rangle).
\end{eqnarray}
Same as Ref. \cite{ControlledTele3}, Alice sends the photons $a_2$
and $b_2$ to Bob, and the photons $a_3$ and $b_3$ to Charlie,
respectively. She performs Bell-state measurements on the photons
$x$ and $a_1$, and $y$ and $b_1$, then the state of the unknown
quantum system $\vert\Phi\rangle_{xy}$ is transferred to the
quantum system composed of the four photons $a_2$, $a_3$, $b_2$,
and $b_3$. When the two agents want to recover the unknown state
$\vert\Phi\rangle_{xy}$, one of them performs two single-photon
measurements on his photons and the other can get the state with
two local unitary operations. In detail, the state of the
composite quantum system composed of the eight photons
$x,y,a_1,a_2,a_3,b_1,b_2$ and $b_3$ can be written as
\begin{eqnarray}
\left\vert\Psi\right\rangle_s&\equiv&\left\vert\Phi\right\rangle_{xy}
\otimes\left\vert\Psi\right\rangle_{a_1a_2a_3}\otimes
\left\vert\Psi\right\rangle_{b_1b_2b_3}\nonumber\\
&=&(a\left\vert00\right\rangle+b\left\vert01\right\rangle+
c\left\vert10\right\rangle+d\left\vert11\right\rangle)_{xy}\nonumber\\
&\otimes&\frac{1}{\sqrt{2}}(\left\vert000\right\rangle+\left\vert111\right\rangle)_{a_1a_2a_3}
\nonumber\\&\otimes&\frac{1}{\sqrt{2}}(\left\vert000\right\rangle+\left\vert
111\right\rangle)_{b_1b_2b_3}.
\end{eqnarray}
When Alice performs the Bell-state measurement on the photons $x$
and $a_1$, they are randomly in one of the four Bell states
\begin{eqnarray}
\left\vert \phi ^{\pm}\right\rangle =\frac{1}{\sqrt{2}}(\left\vert
0\right\rangle\left\vert 0\right\rangle\pm\left\vert
1\right\rangle\left\vert 1\right\rangle), \label{EPR12}\\
\left\vert \psi ^{\pm}\right\rangle =\frac{1}{\sqrt{2}}(\left\vert
0\right\rangle\left\vert 1\right\rangle \pm\left\vert
1\right\rangle\left\vert 0\right\rangle). \label{EPR34}
\end{eqnarray}
So does the quantum system composed of the photons $y$ and $b_1$.

As an example for demonstrating the principle of this QSTS scheme,
we assume that the outcomes obtained by Alice are
$\vert\phi^+\rangle_{xa_1}$ and $\vert\phi^+\rangle_{yb_1}$. Then
the retained four photons are in the state $\left\vert
\Psi\right\rangle_{r}$. Here
\begin{eqnarray}
\left\vert \Psi\right\rangle_{r} &=& a\vert
00\rangle_{a_2b_2}\vert 00\rangle_{a_3b_3} + b \vert
01\rangle_{a_2b_2}\vert
01\rangle_{a_3b_3} \nonumber\\
&+& c\vert 10\rangle_{a_2b_2}\vert 10\rangle_{a_3b_3} + d\vert
11\rangle_{a_2b_2}\vert 11\rangle_{a_3b_3}.
\end{eqnarray}
If Bob performs the product measurement $\sigma_x\otimes\sigma_x$
on the photons $a_2$ and $b_2$, the state $\left\vert
\Psi\right\rangle_{r} $ can be written as
\begin{widetext}
\begin{center}
\begin{eqnarray}
\left\vert \Psi\right\rangle_{r} = \frac{1}{2}&[&\vert
+x\rangle_{a_2} \vert +x\rangle_{b_2} (a\vert 00\rangle_{a_3b_3} +
b \vert 01\rangle_{a_3b_3} + c\vert 10\rangle_{a_3b_3} + d\vert
11\rangle_{a_3b_3})\nonumber\\
&+& \vert +x\rangle_{a_2} \vert -x\rangle_{b_2} (a\vert
00\rangle_{a_3b_3} - b \vert 01\rangle_{a_3b_3} + c\vert
10\rangle_{a_3b_3} - d\vert 11\rangle_{a_3b_3})\nonumber\\
&+& \vert -x\rangle_{a_2} \vert +x\rangle_{b_2} (a\vert
00\rangle_{a_3b_3} + b \vert 01\rangle_{a_3b_3} - c\vert
10\rangle_{a_3b_3} - d\vert 11\rangle_{a_3b_3})\nonumber\\
&+& \vert -x\rangle_{a_2} \vert -x\rangle_{b_2} (a\vert
00\rangle_{a_3b_3} - b \vert 01\rangle_{a_3b_3} - c\vert
10\rangle_{a_3b_3} + d\vert 11\rangle_{a_3b_3})].
\end{eqnarray}
\end{center}
\end{widetext}
That is, Charlie can recover the unknown state
$\vert\Phi\rangle_{xy}$ with two local unitary operations if Bob
and Charlie act in concert. When the outcomes of the product
measurement $\sigma_x\otimes\sigma_x$ done by Bob are $\vert
+x\rangle\vert +x\rangle$, $\vert +x\rangle\vert -x\rangle$,
$\vert -x\rangle\vert +x\rangle$, and $\vert -x\rangle\vert
-x\rangle$, Charlie should perform the unitary operations
$U_0\otimes U_0$, $U_0\otimes U_1$, $U_1\otimes U_0$, and
$U_1\otimes U_1$, respectively on the photons $a_3$ and $b_3$ to
transfer them to the unknown quantum system. Here
\begin{eqnarray}
U_{0}=\left\vert 0\right\rangle \left\langle 0\right\vert
+\left\vert 1\right\rangle \left\langle 1\right\vert, \,\,\,\,\,
U_{1}=\left\vert 0\right\rangle \left\langle 0\right\vert
-\left\vert 1\right\rangle \left\langle 1\right\vert. \label{U1}
\end{eqnarray}

For the other case, the relation between the measurement results
and the final state and the operations needed to reconstruct the
original state $\vert \Phi\rangle_{xy}$ is shown in Table I.
Similar to Ref. \cite{ControlledTele3}, we define $V$ as the bit
value of the Bell state, i.e., $V_{\vert\phi^{\pm}\rangle}\equiv
0$, $V_{\vert\psi^{\pm}\rangle}\equiv 1$. That is, the bit value
$V=0$ if the states of the two particles are parallel, otherwise
$V=1$. $P$ denotes the parity of the Bell-basis measurement and
the single-particle measurement $\sigma_x$, i.e.,
\begin{eqnarray}
P_{\vert\phi^{\pm}\rangle}\equiv\pm,\,\,\,\,
P_{\vert\psi^{\pm}\rangle}\equiv\pm,\,\,\,\, P_{\vert\pm
x\rangle}\equiv\pm,
\end{eqnarray}
and
\begin{eqnarray}
U_{2}=\left\vert 1\right\rangle \left\langle 0\right\vert
+\left\vert 0\right\rangle \left\langle 1\right\vert, \,\,\,\,\,\,
U_{3}=\left\vert 0\right\rangle \left\langle 1\right\vert
-\left\vert 1\right\rangle \left\langle 0\right\vert. \label{U2}
\end{eqnarray}

Different from Ref. \cite{ControlledTele3}, it is unnecessary for
Alice to take a Hadamard operation on each photon in the GHZ state
in this QSTS scheme. Also, the agents need not take a
controlled-NOT (CNOT) gate on the two photons retained for
transferring them to the originally unknown two-qubit state $\vert
\Phi\rangle_{xy}$, only two local unitary operations, when the
parties cooperate. Moreover, the agents need only two
single-photon measurements, not multipartite entanglement
measurements, which makes this QSTS scheme more convenient than
that in Ref. \cite{ControlledTele3}. Also, its efficiency for
qubits $\eta_q$ is 100\% as  all the quantum resources are useful
in theory.

\begin{widetext}
\begin{center}
\begin{table}
\caption{The relation between the local unitary operations and the
results $R_{xa_1}$, $R_{yb_1}$, $R_{a_2}$ and $R_{b_2}$.
$\Phi_{a_3b_3}$ is the state of the two particles hold in the hand
of Charlie after all the measurements are done by Alice and Bob;
$U_C$ are the local unitary operations with which Charlie can
reconstruct the unknown state $\vert \Phi\rangle_{xy}$. }
\begin{tabular}{ccccccc|cccccc}\hline
$V_{xa_1}$  & & $V_{yb_1}$& & $P_{xa_1}\otimes P_{a_2}$ & &
$P_{yb_1}\otimes P_{b_2}$ & & & & $\Phi_{a_3b_3}$ & & $U_C$\\\hline
 0 & & 0 & & $+$ & & $+$ & &  & &
 $a\vert00\rangle +b\vert01\rangle +
  c\vert10\rangle +d\vert11\rangle$ &          & $U_0\otimes U_0 $ \\
0 & & 0 & & $+$ & & $-$ & &  & &
 $a\vert00\rangle -b\vert01\rangle +
  c\vert10\rangle -d\vert11\rangle$ &          & $U_0\otimes U_1 $ \\
  0 & & 0 & & $-$ & & $+$ & &  & &
 $a\vert00\rangle +b\vert01\rangle -
  c\vert10\rangle -d\vert11\rangle$ &          & $U_1\otimes U_0 $ \\
  0 & & 0 & & $-$ & & $-$ & &  & &
 $a\vert00\rangle -b\vert01\rangle -
  c\vert10\rangle +d\vert11\rangle$ &          & $U_1\otimes U_1 $ \\
0 & & 1 & & $+$ & & $+$ & &  & &
 $a\vert01\rangle +b\vert00\rangle +
  c\vert11\rangle +d\vert10\rangle$ &          & $U_0\otimes U_2 $ \\
0 & & 1 & & $+$ & & $-$ & &  & &
 $a\vert01\rangle -b\vert00\rangle +
  c\vert11\rangle -d\vert10\rangle$ &          & $U_0\otimes U_3 $ \\
0 & & 1 & & $-$ & & $+$ & &  & &
 $a\vert01\rangle +b\vert00\rangle -
  c\vert11\rangle -d\vert10\rangle$ &          & $U_1\otimes U_2 $ \\
0 & & 1 & & $-$ & & $-$ & &  & &
 $a\vert01\rangle -b\vert00\rangle -
  c\vert11\rangle +d\vert10\rangle$ &          & $U_1\otimes U_3 $ \\
1 & & 0 & & $+$ & & $+$ & &  & &
 $a\vert10\rangle +b\vert11\rangle +
  c\vert00\rangle +d\vert01\rangle$ &          & $U_2\otimes U_0 $ \\
1 & & 0 & & $+$ & & $-$ & &  & &
 $a\vert10\rangle -b\vert11\rangle +
  c\vert00\rangle -d\vert01\rangle$ &          & $U_2\otimes U_1 $ \\
1 & & 0 & & $-$ & & $+$ & &  & &
 $a\vert10\rangle +b\vert11\rangle -
  c\vert00\rangle -d\vert01\rangle$ &          & $U_3\otimes U_0 $ \\
1 & & 0 & & $-$ & & $-$ & &  & &
 $a\vert10\rangle -b\vert11\rangle -
  c\vert00\rangle +d\vert01\rangle$ &          & $U_3\otimes U_1 $ \\
1 & & 1 & & $+$ & & $+$ & &  & &
 $a\vert11\rangle +b\vert10\rangle +
  c\vert01\rangle +d\vert00\rangle$ &          & $U_2\otimes U_2 $ \\
1 & & 1 & & $+$ & & $-$ & &  & &
 $a\vert11\rangle -b\vert10\rangle +
  c\vert01\rangle -d\vert00\rangle$ &          & $U_2\otimes U_3 $ \\
1 & & 1 & & $-$ & & $+$ & &  & &
 $a\vert11\rangle +b\vert10\rangle -
  c\vert01\rangle -d\vert00\rangle$ &          & $U_3\otimes U_2 $ \\
1 & & 1 & & $-$ & & $-$ & &  & &
 $a\vert11\rangle -b\vert10\rangle -
  c\vert01\rangle +d\vert00\rangle$ &          & $U_3\otimes U_3 $ \\
\hline
\end{tabular}
\end{table}
\end{center}
\end{widetext}

\subsection{QSTS of an arbitrary two-qubit state with $n+1$ agents}

It is straightforward to generalize this QSTS scheme to the case
with n+1 agents, similar to Ref. \cite{ControlledTele3}. For this
multi-party QSTS, Alice will prepare two $(n+2)$-photon GHZ states
and share them with the n+1 agents securely first. Then the
composite state of the system is
\begin{eqnarray}
\left\vert\Psi\right\rangle_S&\equiv&\left\vert\Phi\right\rangle_{xy}
\otimes\left\vert\Psi\right\rangle_{s_1}\otimes\left\vert\Psi\right\rangle_{s_2}\nonumber\\
&=&(a\left\vert 00\right\rangle+b\left\vert 01\right\rangle+
c\left\vert 10\right\rangle+d\left\vert 11\right\rangle)_{xy}\nonumber\\
&\otimes&\frac{1}{\sqrt{2}}(\prod_{i=1}^{n+2}\left\vert0\right\rangle_{a_i}
+\prod_{i=1}^{n+2}\left\vert 1\right\rangle_{a_i})\nonumber\\
&\otimes&\frac{1}{\sqrt{2}}(\prod_{i=1}^{n+2}\left\vert0\right\rangle_{b_i}
+\prod_{i=1}^{n+2}\left\vert 1\right\rangle_{b_i})
\end{eqnarray}
After Alice takes the Bell-state measurements on the particles $x$
and $a_1$, and $y$ and $b_1$, respectively, the state of the
subsystem (without being normalized) becomes
\begin{eqnarray}
\Psi_{sub}&=&
\alpha\prod_{i=2}^{n+2}\left\vert0\right\rangle_{a_i}
\prod_{i=2}^{n+2}\left\vert0\right\rangle_{b_i}+\beta\prod_{i=2}^{n+2}\left\vert0\right\rangle_{a_i}
\prod_{i=2}^{n+2}\left\vert1\right\rangle_{b_i}\nonumber\\
&+&\gamma\prod_{i=2}^{n+2}\left\vert1\right\rangle_{a_i}
\prod_{i=2}^{n+2}\left\vert0\right\rangle_{b_i}+\delta\prod_{i=2}^{n+2}\left\vert1\right\rangle_{a_i}
\prod_{i=2}^{n+2}\left\vert1\right\rangle_{b_i}
\end{eqnarray}
The relation between the parameters $\alpha, \beta, \gamma,
\delta$ and the results $R_{xa_1}$ and $R_{yb_1}$ is shown in
Table II.

When the agents want to reconstruct the original state
$\vert\Phi\rangle_{xy}$, $n$ agents perform the single-photon
product measurement $\sigma_x\otimes\sigma_x$ on their photons
independently. Let us assume that the agents who measure the
photons are Bob$_i$ $(i=2,3,..,n+1)$ and the one who does not
measure his photons is Charlie. That is, the Bob$_i$ act as the
controllers for recovering the unknown state. The measurements
done by them can be express by the formula $M$\\
\begin{eqnarray}
M\equiv[(\langle +x\vert)^{n-t}(\langle -x\vert)^t]_a
\otimes[(\langle+x\vert)^{n-q}(\langle-x\vert)^q]_b.
\end{eqnarray}
Here $[(\langle +x\vert)^{n-t}(\langle -x\vert)^t]_a$ is the
measurement operation related to the state of $a_i$, and
$[(\langle +x\vert)^{n-q}(\langle -x\vert)^q]_b$ is related to
$b_i$. t and q are the numbers of the controllers who obtain the
result $\langle-x\vert$ when they measure the particle $a_i$ and
$b_i$, respectively. After the measurements done by Alice and the
$n$ controllers, the final state $ \Phi_{a_{n+2}b_{n+2}}$ can be
gained by means of performing the operation $M$ on the state
$\Psi_{sub}$,
\begin{widetext}
\begin{eqnarray}
\Phi_{a_{n+2}b_{n+2}} &=&
M(\alpha\prod_{i=2}^{n+2}\vert0\rangle_{a_i}\prod_{i=2}^{n+2}\vert0\rangle_{b_i}
  +\beta\prod_{i=2}^{n+2}\vert0\rangle_{a_i}\prod_{i=2}^{n+2}\vert1\rangle_{b_i}
  +\gamma\prod_{i=2}^{n+2}\vert1\rangle_{a_i}\prod_{i=2}^{n+2}\vert0\rangle_{b_i}
 +\delta\prod_{i=2}^{n+2}\vert1\rangle_{a_i}\prod_{i=2}^{n+2}\vert1\rangle_{b_i})\nonumber\\
&=& \alpha\vert0\rangle_{a_{n+2}}\vert0\rangle_{b_{n+2}}+
(-1)^{q}\beta\vert0\rangle_{a_{n+2}}\vert1\rangle_{b_{n+2}}+
(-1)^{t}\gamma\vert1\rangle_{a_{n+2}}\vert0\rangle_{b_{n+2}}+
(-1)^{q+t}\delta\vert1\rangle_{a_{n+2}}\vert1\rangle_{b_{n+2}}.
\end{eqnarray}
\end{widetext}
We define
\begin{eqnarray}
P_1=P_{xa_1}\otimes\prod_{i=2}^{n+1} P_{a_i},\,\,\,\,
P_2=P_{yb_1}\otimes\prod_{i=2}^{n+1} P_{b_i},
\end{eqnarray}
where $P_{xa_1}$ and $P_{yb_1}$ are the results of Bell-basis
measurements done by Alice, and $P_{a_i}$ and $P_{b_i}$ are the
results of single-particle measurements done by the $n$
controllers, respectively.

\begin{table}[!h]
\label{table1} \caption{The relation between the values of
$\alpha, \beta, \gamma, \delta$ and the results of Bell-basis
measurement
 $R_{xa_1}$ and $R_{yb_1}$ }
\begin{tabular}{ccccccc|ccccccccccccc}\hline
$V_{xa_1}$  & & $V_{yb_1}$& & $P_{xa_1}$ & & $P_{yb_1}$ & & & &
$\alpha$ &&& $\beta$ &&& $\gamma$ &&& $\delta$ \\\hline
 0 & & 0 & & $+$ & & $+$ & &  & &
 $+a$ &&& $+b$ &&& $+c$ &&& $+d$ \\
 0 & & 0 & & $+$ & & $-$ & &  & &
 $+a$ &&& $-b$ &&& $+c$ &&& $-d$ \\
 0 & & 0 & & $-$ & & $+$ & &  & &
 $+a$ &&& $+b$ &&& $-c$ &&& $-d$ \\
 0 & & 0 & & $-$ & & $-$ & &  & &
 $+a$ &&& $-b$ &&& $-c$ &&& $+d$ \\
 0 & & 1 & & $+$ & & $+$ & &  & &
 $+b$ &&& $+a$ &&& $+d$ &&& $+c$ \\
 0 & & 1 & & $+$ & & $-$ & &  & &
 $-b$ &&& $+a$ &&& $-d$ &&& $+c$ \\
 0 & & 1& & $-$ & & $+$ & &  & &
 $+b$ &&& $+a$ &&& $-d$ &&& $-c$ \\
 0 & & 1 & & $-$ & & $-$ & &  & &
 $-b$ &&& $+a$ &&& $+d$ &&& $-c$ \\
 1 & & 0 & & $+$ & & $+$ & &  & &
 $+c$ &&& $+d$ &&& $+a$ &&& $+b$ \\
 1 & & 0 & & $+$ & & $-$ & &  & &
 $+c$ &&& $-d$ &&& $+a$ &&& $-b$ \\
 1 & & 0 & & $-$ & & $+$ & &  & &
 $-c$ &&& $-d$ &&& $+a$ &&& $+b$ \\
 1 & & 0 & & $-$ & & $-$ & &  & &
 $-c$ &&& $+d$ &&& $+a$ &&& $-b$ \\
 1 & & 1 & & $+$ & & $+$ & &  & &
 $+d$ &&& $+c$ &&& $+b$ &&& $+a$ \\
 1 & & 1 & & $+$ & & $-$ & &  & &
 $-d$ &&& $+c$ &&& $-b$ &&& $+a$ \\
 1 & & 1 & & $-$ & & $+$ & &  & &
 $-d$ &&& $-c$ &&& $+b$ &&& $+a$ \\
 1 & & 1 & & $-$ & & $-$ & &  & &
  $+d$ &&& $-c$ &&& $-b$ &&& $+a$ \\\hline
\end{tabular}
\end{table}

The relation between the results of the measurements and the local
operations with which Charlie reconstructs the original state when
he cooperate with Bob$_i$ is as the same as that in Table I with a
little modification. That is, $P_{xa_1}\otimes P_{a_2}$,
$P_{yb_1}\otimes P_{b_2}$, and $\Phi_{a_3b_3}$ are replaced by
$P_1$, $P_2$, and $\Phi_{a_{n+2}b_{n+2}}$, respectively.

Same as the case with two agents, this multiparty QSTS does not
require the agents to do Bell-state measurements on their photons.
Whether the number of the controllers is odd or even, the last
agent Charlie need only take two local unitary operations on the
two photons for reconstructing the unknown two-qubit state.

\subsection{QSTS of an arbitrary $m$-qubit state with $n+1$ agents}

An $m$-qubit state can be described as
\begin{eqnarray}
\vert \Phi\rangle_{u}=\sum_{ij\ldots k}a_{\underbrace{ij\ldots
k}_m}\vert \underbrace{ij\ldots k}_m\rangle_{x_1x_2\ldots
x_m},\label{unknownstate2}
\end{eqnarray}
where $i, j, \ldots, k \in \{0,1\}$, and $x_1, x_2, \ldots$, and $
x_m$ are the $m$ particles in the unknown state. For sharing this
$m$-qubit state, Alice will prepare $m$ ($n+2$)-photon GHZ states
and share them with the $n+1$ agents, say Bob$_j$ ($j=1,2,\ldots
n$) and Charlie. Then the state of the composite quantum system
can be written as
\begin{eqnarray}
\left\vert\Psi\right\rangle&\equiv& (\sum_{ij\ldots
k}a_{\underbrace{ij\ldots k}_m}\vert \underbrace{ij\ldots
k}_m\rangle_{x_1x_2\ldots
x_m})\nonumber\\
&\otimes&
\prod_{i=1}^{m}[\frac{1}{\sqrt{2}}(\prod_{j=0}^{n+1}\left\vert0\right\rangle_{b_{ij}}
+\prod_{j=0}^{n+1}\left\vert
1\right\rangle_{b_{ij}})],
\end{eqnarray}
where the particles $b_{ij}$ ($i=1,2,\ldots, m$) are hold in the
hand of Bob$_{j}$ ($j=1,2,\ldots, n$), $b_{i0}$ are the particles
of Alice's, and $b_{in+1}$ are controlled by Charlie. Similar to
Ref. \cite{Yangcp}, after the Bell-state measurements on the
particles $x_i$ and $b_{i0}$ done by Alice ($i=1,2,\ldots, m$),
the unknown $m$-qubit state will be transferred to the quantum
system composed of the photons kept by the agents. Each of Bob$_j$
performs the product measurement $\sigma^1_x \otimes \sigma^2_x
\otimes \ldots \otimes\sigma^m_x$ on his $m$ photons $b_{ij}$
($i=1,2,\ldots, m$), then the state of the photons kept by Charlie
becomes $(U_1 ^{'-1} \otimes \ldots \otimes U_i ^{'-1}\otimes
\ldots \otimes U_m ^{'-1})\vert \Phi\rangle_{u}$. Here $U_i
^{'-1}\otimes U'_i=I$, and the relation between the results of the
measurements and the local unitary operations $U'_i$ is shown in
Table III. The notation $V_i$ is the bit value of the Bell-state
measurement on the particles $x_i$ and $b_{i0}$, and $P_i$ is
defined as
\begin{eqnarray}
P_i=P_{x_ib_{i0}}\otimes\prod_{k=1}^{n} P_{b_{ij}}.
\end{eqnarray}
That is, if $V_i=0$ and $P_i=+$, Charlie performs the unitary
operation $U_0$ on the $i$-th photon for reconstructing the
unknown state $\vert \Phi\rangle_u$.

\begin{table}[!h]
\caption{The relation between the values of $V_i, P_i$ and the
local unitary operation $U_i$. }
\begin{tabular}{c|cccccccc}\hline
$V_i$  &  & 0 &  & 0 & & 1 & & 1 \\\hline
$P_i$  & &  $+$ & & $-$ & & $+$ & & $-$ \\\hline
$U'_i$ & &  $U_0$  & &  $U_1$   & &  $U_2$  & &  $U_3$\\\hline
\end{tabular}
\label{table3}
\end{table}

\section{Discussion and Summary}

In fact, every one of the multiparty quantum state sharing schemes
can be used for multiparty-controlled teleportation by means that
some of the agents act as the controllers. That is, this
multiparty QSTS scheme can also used for completing the task in
Ref. \cite{ControlledTele3} efficiently. For sharing of
multipartite entanglement, the parties should resort to either
multipartite entanglement quantum resources acting as the quantum
channel or multipartite entanglement measurements. At present,
both of them are not easy to be implemented
\cite{Mentanglement1,Mentanglement2,Mentanglement3}. With the
development of technology, they may be feasible in the future.
Compared with the scheme in Ref. \cite{dengmQSTS}, this multiparty
QSTS scheme is more feasible when the unknown quantum system is
$m$-qubit one and the number of the agents is $n$ ($m,n>2$) as the
sender Alice should perform two $(\frac{n\times m}{2}+1)$-photon
GHZ-state measurements in the former. Different from the Ref.
\cite{ControlledTele3}, each of the controllers needs only to take
a product measurement $\sigma^1_x \otimes \sigma^2_x \otimes
\ldots \otimes\sigma^m_x$ on his photons, no multipartite
entanglement measurements, and the last agent can reconstruct the
unknown $m$-qubit state with only $m$ local unitary operations,
not CNOT gates, when he cooperates with the controllers, which
makes this QSTS scheme more convenient than that in the former.

From Table \ref{table3}, one can see that the agent Charlie can
reconstruct the original unknown state with the probability 100\%
in principle if he cooperate with all the other agents. However,
he will only has the probability $\frac{1}{2^m}$ to get the
correct result if one of the other agents does not agree to
cooperate as Charlie has only half of the chance to choose the
correct operation for each qubit according to the information
published by Alice and the other $n$-1 agents. That is, this
multiparty QSTS scheme is a ($n$, $n$) threshold one, same as that
in Ref. \cite{KKI}. On the other hand, as almost all the quantum
source (except for the instances chosen for eavesdropping check)
can be used to carry the quantum information if the agents act in
concert, the intrinsic efficiency for qubits $\eta_q$  in this
scheme approaches 100\%, as same as those in the schemes
\cite{teleportation,TT1,TT2,ShiGuo,Lee,YanFLteleportation1,ControlledTele2,ControlledTele3,Yangcp}
for quantum teleportation and controlled teleportation. Here
\cite{Cabello}
\begin{eqnarray}
\eta_q=\frac{q_u}{q_t},
\end{eqnarray}
where $q_u$ is the number of the useful qubits in QSTS and $q_t$
is the number of transmitted qubits. In our scheme for sharing
$m$-qubit quantum information with $n$+1 agents, $q_u=q_t=(n+1)m$
(Here we exploited the definition of the intrinsic efficiency for
qubits introduced by Cabello for quantum cryptography
\cite{Cabello}). The total efficiency for this scheme can be
calculated as following.
\begin{eqnarray}
\eta_t=\frac{q_u}{q_t+b_t},
\end{eqnarray}
where $b_t$ is the number of the classical bits exchanged.
$b_t=2m+nm=(n+2)m$. That is
$\eta_t=\frac{(n+1)m}{(n+1)m+(n+2)m}=\frac{n+1}{2n+3}$, the
maximal value for QSTS.

In summary, we present a multiparty QSTS scheme for sharing an
arbitrary $m$-qubit state with $m$ GHZ states, removing some ideas
in the symmetric multiparty-controlled teleportation
\cite{ControlledTele3}. It is an optimal one with GHZ states used
as quantum channel for sharing an arbitrary $m$-qubit state as one
of the parties needs only to perform two-photon Bell-state
measurements and the others use single-photon product
measurements. Moreover,  its efficiency for qubits is 100\%, the
maximal value as all the quantum resources are useful in theory,
and each of the agents can act as the receiver. Simultaneously,
this multiparty QSTS scheme can be used for controlled
teleportation efficiently.

\section*{ACKNOWLEDGEMENT}
This work is supported by the National Natural Science Foundation
of China under Grant Nos. 10447106, 10435020, 10254002, A0325401
and 10374010, and Beijing Education Committee under Grant No.
XK100270454.

\end{document}